\documentclass[manuscript=article, journal=jctcce]{achemso}

\usepackage{bm}
\usepackage{quantikz}
\usepackage{natbib}
\usepackage{subcaption}
\setkeys{acs}{articletitle=true}

\usepackage{hyperref}

\usepackage{epstopdf}
\AppendGraphicsExtensions{.tiff}
\author{James Brown}
\email{james.harry.brown@gmail.com}
\affiliation{%
 Good Chemistry Company, 200-1285 West Pender Street Vancouver, BC, V6E 4B1, Canada\\
}%

\title{ Calculating potential energy surfaces with quantum computers by measuring only the density along adiabatic transitions.}

\date{\today}
\begin{document}
\begin{abstract}
We show that chemically-accurate potential energy surfaces (PESs) can be generated from quantum computers by measuring only the density along an adiabatic transition between different molecular geometries. In lieu of using phase estimation, the energy is evaluated by performing line-integration using the inverted real-space Time-Dependant Density Functional Theory Kohn-Sham potential obtained from the geometry-varying densities of the full wavefunction. The accuracy of this method depends on the validity of the adiabatic evolution itself and the potential inversion process (which is theoretically exact but can be numerically unstable), whereas total evolution time is the defining factor for the precision of phase estimation.  We examine the method with a one-dimensional system of two electrons for both the ground and first triplet state in first quantization, as well as the ground state of three- and four- electron systems in second quantization. It is shown that few accurate measurements can be utilized to obtain chemical accuracy across the full potential energy curve, with shorter propagation time than may be required using phase estimation for a similar accuracy. We also show that an accurate potential energy curve can be calculated by making many imprecise density measurements (using few shots) along the time evolution and smoothing the resulting density evolution. Finally, it is important to note that the method is able to classically provide a check of its own accuracy by comparing the density resulting from a time-independent Kohn-Sham calculation using the inverted potential, with the measured density. This can be used to determine whether longer adiabatic evolution times are required to satisfy the adiabatic theorem.

\end{abstract}

\maketitle

\mciteErrorOnUnknownfalse

\section{\label{sec:level1}Introduction}
Quantum computing has the potential to give us a scalable and efficient method for quantum chemistry\cite{Bauer2020, Babush2023, Cao2019, Kassal2011, Whitfield2014} and has proven advantage for calculating Time-Dependant Density Functional Theory (TDDFT) potentials\cite{Whitfield2014} and probable advantage in other domains (such as electron dynamics \cite{Babush2023}). Although, exponential advantage for ground state calculations of quantum systems is not expected \cite{Lee2023}, there are two distinct methods that could be used to prepare ground states efficiently for most systems. These are 1) adiabatic state preparation\cite{Sugisaki2022} (ASP) and 2) algorithms that implement filters on the energy eigenbasis such as Quantum Signal Processing (QSP)\cite{Pathak2023}. The standard version of quantum phase estimation (QPE)\cite{kitaev1995} is another eigenbasis filtering method and that also returns the energy of the prepared state. The success probability of QPE obtaining the desired state is dependant on the overlap of the desired state with the initial state\cite{Pathak2023, Gratsea2022}. Also, the precision $\epsilon$ of this energy estimation depends on the total time evolution as $\mathcal{O}\left(1/\epsilon\right)$. This means that a time propagation of $\approx 1000 $ a.u. (where a.u. is atomic units) may be necessary to generate submillihartree (a.k.a. chemical) accuracy\cite{Sugisaki2022}.   ASP or QSP does not provide direct access to the energy but can sometimes prepare a state more efficiently than QPE,\cite{Sugisaki2022, Gratsea2022} depending on various factors such as the energy gap between the two lowest states along the adiabatic path\cite{Sugisaki2022}.  ASP also does not significantly depend on the overlap of the initial wavefunction with the desired state\cite{Sugisaki2022}. However, if one wishes to use ASP or QSP to obtain an estimate of the prepared ground state for a single geometry, a method such as QPE or measuring the expectation value of each term in the Hamiltonian\cite{Peruzzo2014} would also be necessary after the ground state has been prepared.

Here we show that when energies of many geometries are desired, such as generating potential energy surfaces (PESs), the use of QPE after ASP/QSP is unnecessary.  After preparing a ground state at a certain geometry (where the ground state is relatively easy to prepare), measuring only the density along the adiabatic time evolution between different external potentials is sufficient to generate energy differences, and therefore PESs. These energy differences are calculated using TDDFT along with a DFT adiabatic line-integral using the inverted exact TDDFT Kohn-Sham (KS) potential.

To obtain the exact KS potential, one needs to solve the inverse Kohn-Sham (iKS) problem. This problem is (in general) numerically unstable and difficult\cite{Brown2020, n2v}, and this is especially true when using standard incomplete Gaussian basis sets where unphysical oscillations often occur,\cite{n2v} but iKS is less problematic when using a grid basis\cite{Chayes1985}. Therefore, to remove some of the difficulties in applying the proposed technique, we mainly focus on utilizing a grid basis. To obtain the KS potential, one needs access to the exact density of the system. However, obtaining exact densities on a classical computer using a grid basis for more than single electron systems (which is what DFT is) is impractical due to the very large basis set required. However, this memory bottleneck is not an issue for quantum computers and the grid basis is one of the leading candidates for advantage using future quantum computers\cite{Babush2023}. Therefore, although most of the discussion in this work is limited to 1D systems due to current memory constraints, the techniques can be applied to future quantum computers. To demonstrate the difficulties of using standard basis sets, we include a discussion on the H$_4$ dissociation in a 6-31G(d) and 6-31G(d,p) basis and discuss various problems that arise.

In section \ref{sec:algorithm}, we outline the theory behind the method. In section \ref{sec:examples}, we show that the method is applicable to singlet and triplet states of a first-quantized two-electron, and also second-quantized three- and four-electron systems. In section \ref{sec:pes_gen}, we discuss how one may be able to study quantum dynamics of chemical systems using data from the method introduced in this manuscript by generating full potential energy surfaces. In section \ref{sec:vqpe}, we provide an estimate of where the algorithm presented here may be useful compared to QPE followed by concluding remarks in section \ref{sec:conclude}.

\section{Algorithm\label{sec:algorithm}}
In Kohn-Sham Density Functional Theory (DFT), a fictitious non-interacting system $\Phi\left(\mathbf{r}\right)$ is introduced as a Slater determinant of $N$ one-electron orbitals $\phi_i\left(r\right)$ that has a density equivalent to the full interacting system $\Psi\left(\mathbf{r}\right)$ with $N$ electrons such that,
\begin{equation}\label{eq.dens_equal}
\begin{split}
\rho\left(r\right)=N\int \mathrm{d}r_2...\mathrm{d}r_N \Psi^*\left(r_1, r_2, ..., r_N\right)\Psi\left(r_1, r_2, ..., r_N\right) \\ \equiv \sum_i^N \phi_i^*\left(x_i\right)\phi_i\left(x_i\right)=\rho_{KS}\left(r\right)
\end{split}
\end{equation}
where $\rho\left(r\right)$ is the density, $\mathbf{r}$ refers to all $N$ one-electron coordinates $r_i$, and $x_i$ refers to spin-orbital coordinates that include both the electron spin $\sigma_i$ and $r_i$.

The full interacting system is the ground state eigenfunction of the electronic Hamiltonian,
\begin{equation}\label{eq.full_H}
    H_{R} = \sum_{i}^{N}\left[T_i + v_{ext}\left(r_i;R\right) \right]+ \sum_{i=1}^N\sum_{j>i}^N W\left(r_i, r_j\right)+V_{nn}(R)
\end{equation}
where $W\left(r_i, r_j\right)$ is the interaction potential and $v_{ext}(r_i;R)$ (which depends parametrically on the nuclear coordinates $R$) is the one-particle external potential and $T_i=-\frac{\nabla_i^2}{2}$ is the kinetic energy, and $V_{nn}(R)$ is the nuclear potential energy which does not involve the electron coordinates. The KS orbitals are eigenfunctions of the fictitious one-electron non-interacting system
\begin{equation}\label{eq.KS_H}
    h_R = T_i + v_{ext}\left(r_i;R\right) + v_{KS}\left(r_i\right)+V_{nn}(R)
\end{equation}
 where $v_{KS}\left(r_i\right)$ is the one-particle potential that ensures that Eq. (\ref{eq.dens_equal}) is followed.  For the KS system, the total energy of the system is calculated by using the energy functional
\begin{equation}
    E_{KS} = T\left[\left\{\Phi\right\}\right] + V_{ext}\left[\rho\left(r\right)\right] + F\left[\rho\left(r\right)\right],
\end{equation}
where $T\left[\left\{\Phi\right\}\right]$ is the kinetic energy functional,  $V_{ext}\left[\rho\left(r\right)\right]$ is the external potential functional and $F\left[\rho\left(r\right)\right]$ is the Hartree exchange correlation functional. The KS potential is defined as $v_{KS}=\frac{\partial F}{\partial \rho}$. In references \citenum{vanLeeuwen1995, Liardi2021}, it was shown that energy differences between different external potentials can be calculated by using the line integral formula
\begin{equation}\label{eq.line_int}
    E_{KS}\left(\rho_B\right)-E_{KS}\left(\rho_A\right) = \int \mathrm{d}t \int \mathrm{d} r \,v_{KS}\left(r, t\right)\frac{\partial \rho\left(r,t\right)}{\partial t},
\end{equation}
where $\rho\left(r, t\right)$ is such that the properties of Eqs. \ref{eq.dens_equal}, \ref{eq.full_H} and \ref{eq.KS_H} are satisfied for all time points with external potential $v_{ext}\left(r, t\right)$. That is, the KS and fully interacting systems remain in the ground state, and the densities match. In order to naturally apply this result to quantum computers, we adiabatically time-evolve a system from an initial density $A$ to another density $B$, and measure the density along the transition. Along this adiabatic path, we obtain $v_{KS}(r,t)$ by using the TDDFT potential inversion algorithm of reference \citenum{Brown2020}.

\subsection{Potential inversion}
The potential inversion algorithm of reference \citenum{Brown2020} requires that we have an initial set of orbitals $\phi_i\left(r\right)$ such that Eq. \ref{eq.dens_equal} is satisfied, and the first time derivative of the density of the full system is equal to that of the KS system. Assuming that our starting point is an eigenstate of Eq. \ref{eq.full_H} with density $\rho_A$, we can solve for the ground state orbitals using the method of references \citenum{Baker2020, Gidopoulos2011}. This involves solving
\begin{equation}\label{eq.gen_orbs}
\begin{split}
    T_{\Psi}\left[\rho\right] = \big<\Psi\left(\mathbf{r}\right)\big|\left(T+v_{ext}+v_{KS}\right)\big|\Psi\left(\mathbf{r}\right)\big>-\\ \big<\Phi\left(\mathbf{r}\right)\big|\left(T+v_{ext}+v_{KS}\right)\big|\Phi\left(\mathbf{r}\right)\big>
    \end{split}
\end{equation}
which generates the initial $v_{KS}(r, 0)$ potential and $\phi_i\left(r\right)$ KS orbitals\cite{Baker2020, Gidopoulos2011} for the starting $\rho_A$. A grid-like basis is also required to utilize the TDDFT potential inversion algorithm of reference \citenum{Brown2020}. Therefore, we use the basis set mentioned in references \citenum{Babush2023, Childs2022, Chan2023}, which is hypothesized to be quite efficient for quantum chemistry on quantum computers\cite{Babush2023, Childs2022}. This is a discrete variable representation\cite{Light2000} but more specifically the Fourier Grid Hamiltonian\cite{Marston1989} where the matrix elements for the potential are $V_{ij}=v(x_i)\delta_{ij}$ at equally space grid points $x_{i} = i\Delta x$ for grid spacing $\Delta x$. $v(r)$ for the full system is $v_{ext}(r)$ where for the KS system is $v_{ext}(r)+v_{KS}(r)$. In Fourier space, the kinetic energy matrix elements are also diagonal with elements $k_{ij}=\frac{k_i^2}{2}\delta_{ij}$ with $k_i=i \frac{2\pi}{\Delta x N}$ where $N$ is the number of grid points.

The first term of Eq (\ref{eq.gen_orbs}) requires accessing the expectation value of $\big<\Psi\left(\mathbf{r}\right)\big|\left(T+v_{ext}+v_{KS}\right)\big|\Psi\left(\mathbf{r}\right)\big>$ from the full wavefunction $\Psi$. $\big<\Psi\left(\mathbf{r}\right)\big|\left(T+v_{ext}+v_{KS}\right)\big|\Psi\left(\mathbf{r}\right)\big>$ can be obtained by using a state-preserving quantum counting algorithm\cite{Baker2020, Temme2011} if the state preparation for the ground state is expensive. However, as both $T$ and $v_{ext}$ only includes one-body terms, it can also be fully calculated by measuring the one-body RDM once. The function can be minimized by recalculating the lowest eigenfunctions $\phi_i\left(r_i\right)$ for each candidate $v_{KS}$ classically in $\mathcal{O}\left(N \log N\right)$ time due to the potential being diagonal and $T$ can be diagonalized with the Fast Fourier Transform (FFT)\cite{van1992computational} algorithm. It should also be noted that the Jacobian of Eq \ref{eq.gen_orbs} is easily calculated as $\frac{\mathrm{d}}{\mathrm{d} v_{KS}^{(j)}} = \rho_{j}\left(r\right)-\sum_i \left|\phi_i^{(j)}\left(r\right)\right|^2$ where the superscript $(j)$ indicates the $j^{th}$ grid point of $v_{KS}(r)$ or $\phi_i(r)$.

Most of the numerics for this manuscript are performed using TDDFTinversion\cite{TDDFTinversion}, which is developed using the methodology outlined in reference \citenum{Brown2020}. TDDFTinversion is mainly written in Fortran, but has a Python interface that utilizes f90wrap\cite{f90wrap}. In section \ref{sec:qc_calc}, we simulate the time evolution on a quantum simulator using the discrete clock construction of reference \cite{Watkins2022}, as implemented in Tangelo\cite{Tangelo}. 

\subsection{Adiabatic scheduling}
The success of the adiabatic transition depends on the scheduling function used to propagate the system from an eigenstate of Hamiltonian $A$ to an eigenstate of Hamiltonian $B$. At least the first and second derivatives must be zero at the end-points to ensure proper adiabatic evolution \cite{GuryOdelin2019}. We find that also ensuring the third derivative is zero is beneficial to ensure that the density evolution is smooth enough for practical matters. This becomes useful when attempting to obtain information about the time evolution of the density from the quantum computer. Therefore, we use the simplest polynomial that satisfies the conditions that $S(0)= S'(0)=S''(0)=S'''(0)=S'(1)=S''(1)=S'''(1)=0$ and $S(1)=1$, which is 
\begin{equation}\label{eq.sched}
    -20s^7+70s^6-84s^5+35s^4, \quad s=\frac{t}{T}.
\end{equation}

It should be noted that reference \citenum{Sugisaki2022} empirically found that the scheduling for their use cases which resulted in the shortest time evolution to obtain large overlaps was $\sin(t)$ which does not satisfy the adiabatic evolution conditions\cite{GuryOdelin2019}. The reason for this is that they always use a Hartree-Fock reference state and Fock operator as the initial state and operator respectively.
The Fock operator Hamiltonian contains the terms with occupied orbitals only and
the excited states cannot be sufficiently stabilized under the Fock
operator\cite{Sugisaki2022}. This is not satisfied for any fully interacting to fully interacting transitions. That being said, the $\sin(t)$ scheduling function could be used to obtain the initial ground state from a Hartree-Fock reference if QSP is not utilized. Generating PESs using the technique outlined in this manuscript will be most beneficial when the ground state preparation is efficient for at least one molecular geometry. This will occur when the Hartree-Fock reference is a good approximation to the exact ground state. The Hartree-Fock reference could be efficiently prepared using the technique outlined in reference \citenum{Babush2023}. Further work will examine how improved scheduling can be achieved for fully interacting to fully interacting transitions.

In most instances of ASP, one evolves from $R_A$ to $R_B$ using
\begin{equation}\label{eq.alchemical}
    H(t)=H_{R_A}[1-S(t)]+H_{R_B}S(t),
\end{equation}
but we use 
\begin{equation}\label{eq.chemical}
    H(t)= H_{\left[R_B-R_A\right]S(t)+R_A},
\end{equation}
where $R_A$ defines the geometry at $A$ and $R_B$ defines the geometry of the molecular system at $B$. This adiabatic process means that we can get approximate energies and $v_{KS}$ at geometries along the full time evolution.

In reference \citenum{Yu2022}, they utilize a scheduling of Eq \ref{eq.alchemical} for molecular systems after obtaining an easy to prepare state at geometry $R_A$. They show that there is far less of a chance of energy crossings compared to starting from a Hartree Fock state at each geometry. We would assume that this would continue for our use-case of Eq. \ref{eq.chemical} where breakdown would only occur at conical intersections.

\section{Examples\label{sec:examples}}
Before utilizing a quantum computer simulator, we first validate the method using exact simulation and perfect measurement. The exact evolution/ exact measurement experiments use TDDFTinversion to 1) Generate the ground state eigenfunctions and one-body RDM, 2) scipy\cite{2020SciPy-NMeth} to calculate the initial orbitals and $v_{KS}$ using Eq (\ref{eq.gen_orbs}), and  3) TDDFTinversion to propagate both the full system and generate the KS orbitals along the adiabatic path from the exact second time-derivative of the density. 

\subsection{Exact evolution\label{sec:exact}}
The first system we look at is the frozen core one-dimensional two-electron LiH model system defined in references \citenum{Tempel2009, Kosugi2022}. The system assumes a frozen core Li atom at $-R/2$ and an H atom at $R/2$. The Hamiltonian for the system is,
\begin{equation}
\begin{split}\label{eq:lih_system}
    H_R = \sum_{i=1}^2\Bigg[ -\frac{1}{2}\frac{\mathrm{d}^2}{\mathrm{d}x_i^2}-\frac{1}{\sqrt{(r_i-\frac{R}{2})^2+0.7}}- \\ \frac{1}{\sqrt{(r_i+\frac{R}{2})^2+2.25}}\Bigg] \\ +\frac{1}{\sqrt{(r_1-r_2)^2+c}} + V_{nn}(R),
\end{split}
\end{equation}
with the nuclear potential energy defined as,
\begin{equation}
    V_{nn}(R)=\frac{1}{\sqrt{\left(2.25+0.7-c\right)+R^2}}.
\end{equation}

This system has been used as a test case for analysing potential inversion methods because of the step-like nature of the $V_{KS}$ potential when dissociating. This step is due to the necessity of the density being split between the nuclei due to the Coulomb repulsion\cite{Tempel2009}. The non-interacting KS system would have both electrons on the model H atom due to its higher electron affinity compared to the screened Li atom if the step in the $V_{KS}$ potential was not present.

\begin{figure}
     \centering
     \begin{subfigure}[b]{0.45\textwidth}
         \centering
         \includegraphics[width=\textwidth]{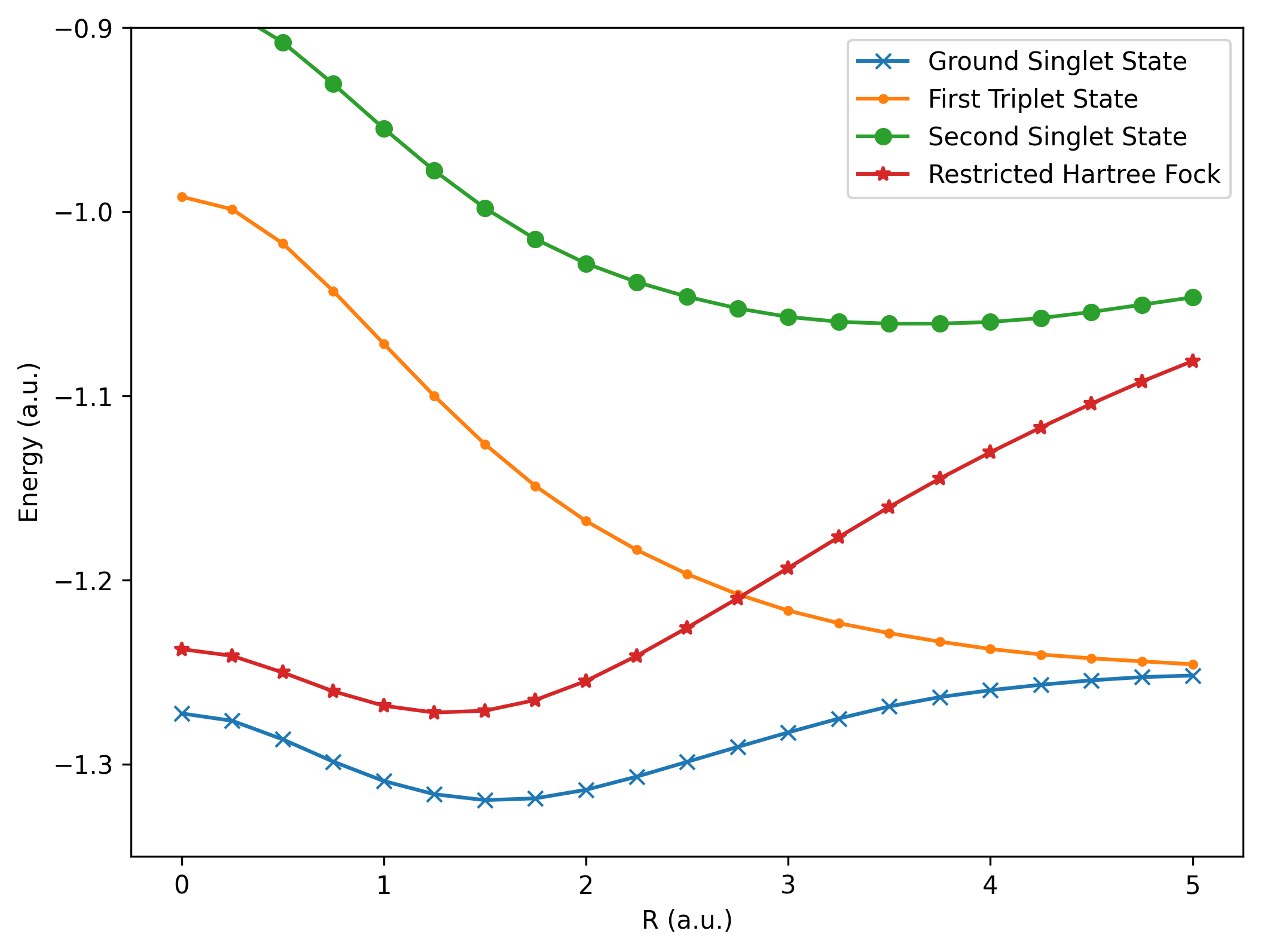}
         \caption{$c=0.6$}
         \label{fig:c06}
     \end{subfigure}
     \hfill
     \begin{subfigure}[b]{0.45\textwidth}
         \centering
         \includegraphics[width=\textwidth]{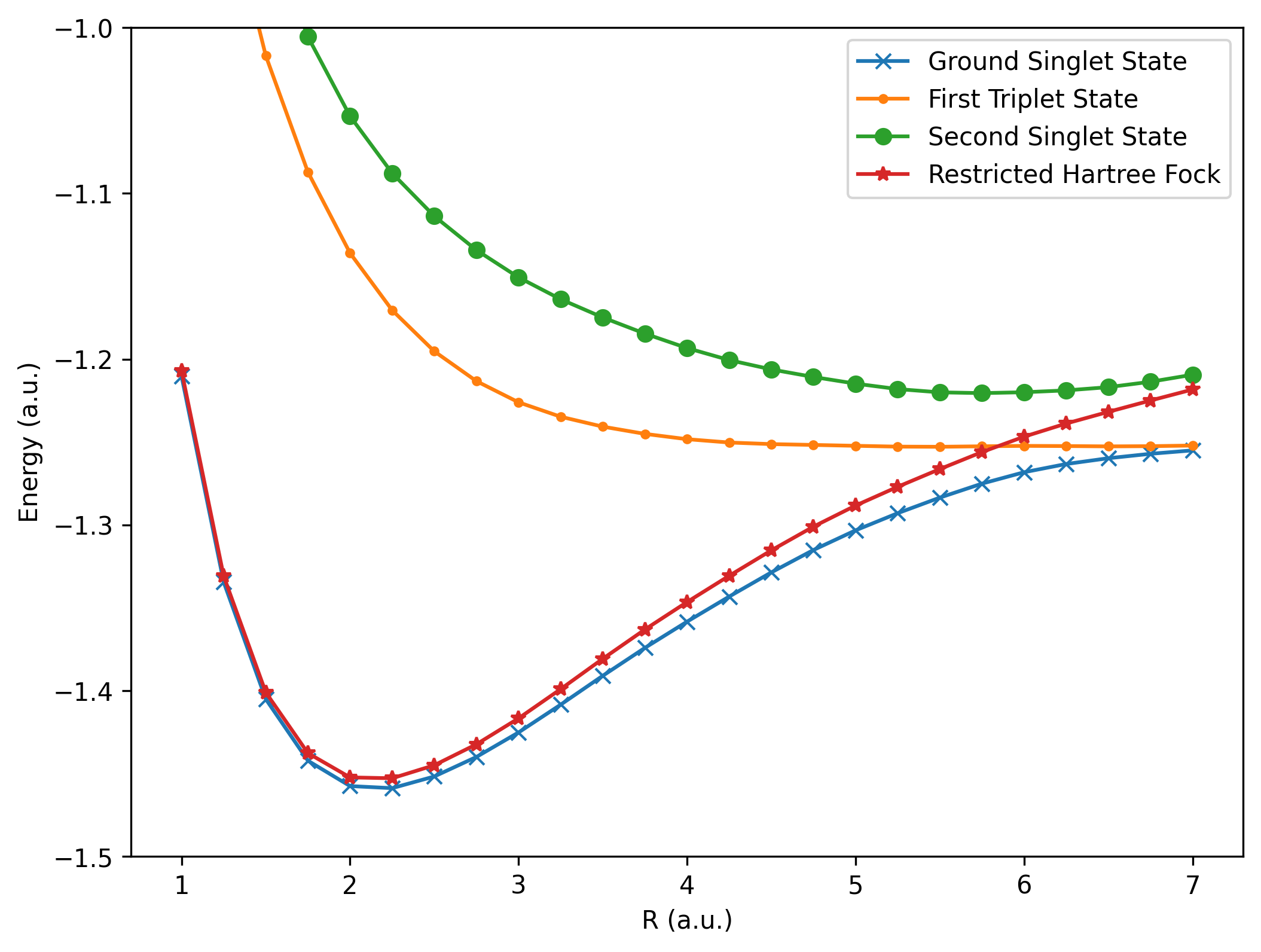}
         \caption{$c=2.8$}
         \label{fig:c28}
     \end{subfigure}
        \caption{The exact energies for the lowest potential energy curves of the model LiH system with Hamiltonian given by Eq.\ref{eq:lih_system} for two different parameters of $c$. A larger value of $c$ pushes out the $R$ at which the avoided crossing occurs and makes it sharper.}
        \label{fig:lih_energies}
\end{figure}

$c$ is an adjustable parameter that controls at which distance (and how sharply) the avoided crossing between the ground and second singlet state occurs. To reproduce the difference in the ionization potentials between the two nuclei, a parameter of $c=0.6$ is used. This produces the potential energy curves shown in figure \ref{fig:c06}. We use $2^5=32$ grid points for each electron (which for this two-electron system is a total simulation size of $2^{10} =1024$ grid points or $10$ qubits in first quantization) with a grid spacing of $0.6$ Bohr. We also show the potential energy curves for $c=2.8$ in figure \ref{fig:c28} but the analysis of the algorithm with this parameter is in the appendix in section \ref{sec:avoided}. This choice of small basis has reasonably good agreement with the original paper, as can be seen by comparing figure \ref{fig:lih_energies} to figure 5 of reference \citenum{Tempel2009}.

For the initial demonstration, we evolve the full system from $R=0.25$ to $R=4.25$ using $c=0.6$. This total evolution time was selected as $144$ a.u. so that the state at the end of the adiabatic evolution had an overlap with the exact ground state (with geometry $R=4.25$) greater than $0.999$. The time-dependant Hamiltonian is then
\begin{equation}
\begin{split}
    H(t) = \sum_{i=1}^2\Bigg[-\frac{1}{2}\frac{\mathrm{d}^2}{\mathrm{d}x_i^2}-\frac{1}{\sqrt{(r_i-\frac{R(t)}{2})^2+0.7}}- \\ \frac{1}{\sqrt{(r_i+\frac{R(t)}{2})^2+2.25}}\Bigg]\\ +\frac{1}{\sqrt{(r_1-r_2)^2+0.6}}+ V_{nn}(R(t))
\end{split}
\end{equation}
where $R(t)=0.25+(4.25-0.25)S(t)$ with $S(t)$ defined in Eq. \ref{eq.sched}.

\begin{figure}
     \centering
     \begin{subfigure}[b]{0.45\textwidth}
         \centering
         \includegraphics[width=\textwidth]{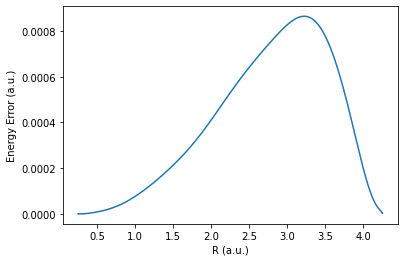}
         \caption{LiH energy error}
         \label{fig:d_err}
     \end{subfigure}
     \hfill
     \begin{subfigure}[b]{0.45\textwidth}
         \centering
         \includegraphics[width=\textwidth]{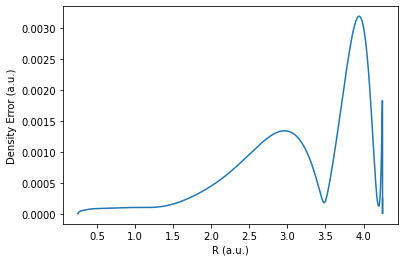}
         \caption{LiH density Error}
         \label{fig:e_err}
     \end{subfigure}
        \caption{a) The energy and b) density error of the adiabatic evolution for the bond stretching of LiH. The calculated values are more accurate at the end points with an energy error of $4.7\times 10^{-7}$ and a density error of $2\times 10^{-4}$}
        \label{fig:Lih_singlet}
\end{figure}

\begin{figure}
     \centering
     \begin{subfigure}[b]{0.45\textwidth}
         \centering
         \includegraphics[width=\textwidth]{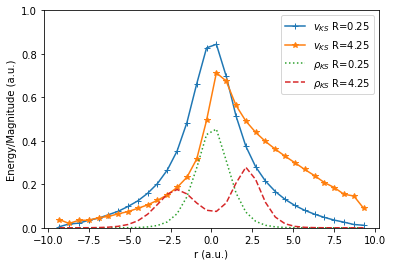}
         \caption{LiH singlet state}
         \label{fig:v_d}
     \end{subfigure}
     \hfill
     \begin{subfigure}[b]{0.45\textwidth}
         \centering
         \includegraphics[width=\textwidth]{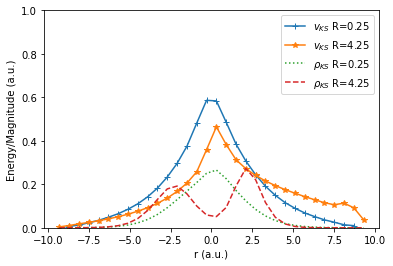}
         \caption{LiH triplet state}
         \label{fig:v_d_t}
     \end{subfigure}
        \caption{The a) singlet and b) triplet density and KS potential at the beginning and end of the time evolution for the Hamiltonian of Eq. \ref{eq:lih_system}}
        \label{fig:Lih_triplet}
\end{figure}

For the Kohn-Sham propagation, it was found that it was necessary to increase the evolution time to $432$ (three times as long to converge the results) due to non-adiabaticity in the KS system. To obtain the $V_{KS}$ potential, we need the second derivative of the density\cite{Brown2020}. For this exact test, the second derivative for intermediate times is taken by interpolating the density from the full system evolution and numerically calculating the second derivative from the interpolation. As we time-evolve the Kohn-Sham system, we can obtain approximations to both the energy and Kohn-Sham potential for each point on the potential energy surface. These are shown in figure \ref{fig:Lih_singlet}. The density error is calculated by taking the instantaneous $v_{KS}(r, t_i)$ for a time $t_i$ and solving the eigenvalue problem (Eq (\ref{eq.KS_H}) to obtain a density $\tilde{\rho}(r)$ and then calculating the norm of the density difference $\sqrt{\int \mathrm{d}r\,\left|\rho(r, t_i)-\tilde{\rho}(r, t_i)\right|^2}$. For the ground state, the KS density is defined as $2\left|\phi_0\right|^2$ since the ground state is doubly occupied in the Slater determinant. As can be seen in figure \ref{fig:Lih_singlet}, the energy error stays below chemical accuracy for the full time evolution, and the density error also remains reasonably accurate. In future work, it may be fruitful to assign phases\cite{Brown2020} to the orbitals such that the time-derivative of the density is zero, and calculate the potential corresponding to a time-independant density evolution. At the end points (i.e. $R=4.25$), when the phases are zero, the energy error is $5 \times 10^{-7}$ while the density error is $2 \times 10^{-4}$, which shows that very high accuracy could be achieved. The line integral of Eq (\ref{eq.line_int}) evaluates to $-0.4130865$ Hartree, which shows that the method can be accurate for the large changes in the density and $v_{KS}$ potential as depicted in figure \ref{fig:v_d}.

\begin{figure}
     \centering
     \begin{subfigure}[b]{0.45\textwidth}
         \centering
         \includegraphics[width=\textwidth]{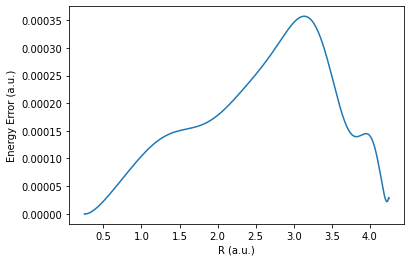}
         \caption{LiH energy error}
         \label{fig:e_err_t}
     \end{subfigure}
     \hfill
     \begin{subfigure}[b]{0.45\textwidth}
         \centering
         \includegraphics[width=\textwidth]{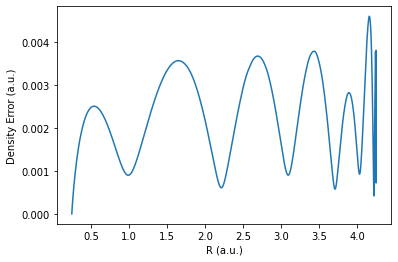}
         \caption{LiH density Error}
         \label{fig:d_err_t}
     \end{subfigure}
        \caption{The a) energy and b) density error of the adiabatic evolution for the first triplet state bond stretching of LiH using the Hamiltonian of Eq. \ref{eq:lih_system}. }
        \label{fig:Lih_singlet_d_e}
\end{figure}

For the triplet state evolution shown in figure \ref{fig:Lih_singlet_d_e}, the density error and energy error are both small during the duration of the adiabatic evolution. The KS density is defined as $\rho_{KS}(r)=\left|\phi_0\right|^2+\left|\phi_1\right|^2$ as the system is in a triplet state. One could use the same occupation (i.e. $2\left|\phi_0\right|^2$) as the ground singlet state but the inverted KS potential becomes much more complicated. From figure \ref{fig:v_d_t}, it can be seen that the triplet $V_{KS}$ potential is of similar shape to the singlet $v_{KS}$ potential at both $R=0.25$ and $R=4.25$ but with different magnitudes. However, the density of the triplet state is very different from the singlet state at $R=0.25$. The densities are similar at $R=4.25$ as is expected for these states at dissociation.

We note that there appears to be a correspondence between the density error and the energy error in figures \ref{fig:Lih_singlet} and \ref{fig:Lih_singlet_d_e}. The correspondence is not one-to-one where high (low) density error means high (low) energy error. But the increased density error from the singlet calculation does correspond to a higher energy error relative to the triplet state. This suggests that one may be able to compare the measured density and the KS density from the inverted potential to provide an estimate of how closely the adiabatic theorem has been satisfied. Further work will study how representative this density discrepancy is and whether it can be utilized to optimize the adiabatic scheduling function. Improvements could be made by using the field of quantum control for which the inherent smoothness herein is beneficial\cite{Li2023}.

It should be made clear that there is path independence when using Eq (\ref{eq.line_int}). One does not need to follow along a geometry change of a molecule to obtain accurate results. The main benefit is that one gets information about the energy at intermediate geometries instead of just the end points. We have tested the method for taking a simple sum of the initial Hamiltonian and final Hamiltonian $H_{R=0.25}(1-S(t))+H_{R=4.25}S(t)$ and similar results are obtained at the end points. In fact, any choice is acceptable as long as the number of $\alpha$ and $\beta$ electrons is held constant. This is the realm of alchemical density functional theory\cite{Krug2022}, where isoelectric species are evolved and is applicable using this method. In fact, the schedule function of $H_{R=0.25}(1-S(t))+H_{R=4.25}S(t)$ performs alchemical density functional theory by morphing the two nuclei system into a four partial charge nuclei system at intermediate times. For cases where avoided crossings occur (which would greatly increase the adiabatic evolution time), it may be beneficial to evolve the system according to Eq. \ref{eq.alchemical} vs Eq. \ref{eq.chemical}. This model LiH system can easily be modified to study this\cite{Tempel2009} by changing the parameter $c$ as shown in figure \ref{fig:lih_energies}.

\subsection{Using density measured from a quantum algorithm\label{sec:qc_calc}}
In order to examine the applicability to quantum computing, we implement the discrete clock time evolution algorithm\cite{Watkins2022} in Tangelo\cite{Tangelo}. In the appendix (\ref{app:dc_grid}), we show that the discrete-clock algorithm works very well for the Fourier grid Hamiltonian basis set\cite{Babush2023} used here. We choose order $M=3$  for the multiproduct component of the algorithm, with a linear combination of $M_j$ Trotter  time-steps of $M_j=1, 2, 6$. The order of $M=3$ is chosen as the density needs to be sampled at multiple times and higher order is only beneficial for longer times between sampling. $M_j$ time-steps of $M_j=1, 2, 6$ limits the times at which the time-dependant Hamiltonian potential need to be generated. The $6$ time-step propagation (i.e. $t_i= 1/12, 3/12, 5/12, 7/12, 9/12, 11/12$) include the times of the $2$ time-step propagation (i.e. $t_i=3/12, 9/12$) so some circuit optimization can be performed. This is explained further in the appendix. The Trotter time-steps are performed using the second-order split-operator method of $e^{-iH(t)t}\approx \prod_i^{M_j} e^{-i\frac{p^2}{2m} \frac{\mathrm{d}t}{2}}e^{-iV(t_i)\mathrm{d}t}e^{-i\frac{p^2}{2m} \frac{\mathrm{d}t}{2}}$. 

The $e^{-i\frac{p^2}{2m}}$ evolution for time $\mathrm{d}t/2$ (denoted $U_T$) is performed by a quantum Fourier transform (QFT) on each electron register and then the diagonal $e^{-i\frac{p^2}{2m}\frac{\mathrm{d}t}{2}}$ is applied using the circuit of reference \citenum{Ollitrault2020}. The potential evolution at time $t$ (denoted $U_V(t)$) are applied by generating the qubit operator that represents the potential at time $t_i$ and using a Trotter step for time $\mathrm{d}t$.  The second order Trotter time-evolution for the Hamiltonian  from time $t$ to $t+\mathrm{d}t$ is $U_2(t,t+\mathrm{d}t)=U_T U_V(t+\mathrm{d}t/2) U_T$. The application of the potential terms is by far the most time consuming step. Better time evolution of the Coulomb interaction term is an active area of research\cite{Kosugi2022}.

In order to have sufficient sampling for the potential inversion, the time used for each multiproduct time-step is $0.4$ with the density sampled after each multiproduct evolution. This time-step provided a good trade-off between measurement requirements and the ability to reconstruct the full density evolution. A shorter time-step would increase the measurement requirements but also increase the accuracy of the reconstructed density evolution. Longer time-steps would decrease the measurement requirements but make it more difficult to reconstruct the exact density evolution. For the purposes of this work, the density is calculated by obtaining the full frequency distribution. But on future fault-tolerant devices, the density should be measured simultaneously using the method outlined in reference \citenum{Babush2023}. For this test, we assume that the one-body RDM has been measured exactly with perfect state preparation. As we are using first-quantization to represent the system, each basis function on a grid is represented as a bitstring $b$ such that the wavefunction is $\sum_b c_b \left|b\right\rangle$, where the anti-symmetric nature is enforced on the coefficients $c_b$ themselves.\cite{Babush2023}  Since we are using $32$ basis functions, we require $5$ qubits to represent each electron or $10$ qubits for this 2-electron system. On a quantum computer, the simplest method one can use to calculate each element of the one-body RDM is using a Hadamard test with the unitary operation that takes $i\rightarrow j$ where $i,j$ are bitstrings. For example, $\left<\psi\big|a_0^{\dagger}a_3\big|\psi\right>$ uses the unitary $X_0 X_1$ in the Hadamard test as $0$ corresponds to bitstring $00000$ and $3$ corresponds to bitstring $11000$. Assuming that either $00000$ or $11000$ was measured in the basis state register. The same unitary could also measure any $00lkm \rightarrow 11lkm$ transition element depending on the measurements of the binary values $l,k,m$. Although this method works, the more efficient method developed in reference \citenum{Babush2023} will probably be what is used in practice to measure the RDM as it has better scaling. That being said, the prefactor of more advanced measurement techniques may overwhelm the optimal scaling as was shown when measuring the density in reference \citenum{Yang2021}.

\subsubsection{Obtaining the 2nd derivative of the density}
As stated previously, in order to propagate the KS system and calculate the corresponding $v_{KS}(r,t)$ using TDDFTinversion, the second derivative of the density needs to be calculated at many points. Directly measuring the second derivative from a quantum computer is impractical due to the non-commuting nature of the terms involved. It is far more efficient to sample the density itself due to the commuting nature of the density elements. Additionally, it is beneficial to compute the first time-derivative of the density, as it can be used to stabilize the TDDFT potential inversion method\cite{Brown2020}. Lastly, the smoothness of the density evolution can be utilized to obtain more accurate data, as shown later in section \ref{sec:many_noisy}. Smoothness has also been shown to be beneficial in the realm of quantum control\cite{Li2023}.

\subsubsection{Using few exact density measurements\label{sec:few_exact}}
\begin{figure}
    \centering
    \includegraphics[width=0.45\textwidth]{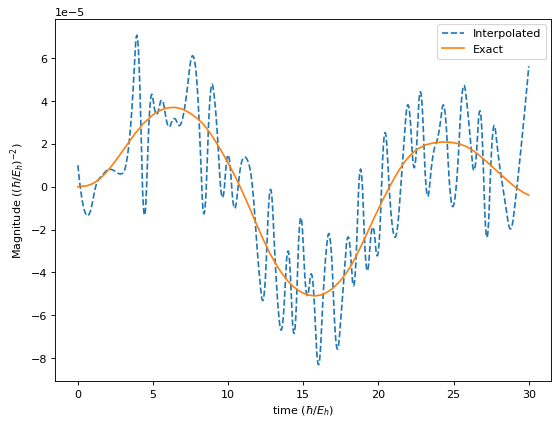}
    \caption{The second derivative of the density from exact numerical simulation vs interpolation and numerical differentiation from 75 sampled points for point $16$ for the Hamiltonian of Eq. \ref{eq:lih_system}.}
    \label{fig:second_der}
\end{figure}

Using the scheduling function outlined in Eq. (\ref{eq.sched}) ensures that the density evolution is smooth. Therefore, we can use the 4th order B-spline of scipy\cite{2020SciPy-NMeth} to interpolate between the 75 measured points with $d t=0.4$ for a total time evolution of 30$a.u$ from the discrete clock simulation for the ground state of the Hamiltonian of Eq \ref{eq:lih_system}. We use a 4th order B-spline to ensure that the second derivative of the interpolated function is continuous and smooth. The difference between the second derivative of the interpolated values and the exact second derivative for grid-point 16 is shown in figure \ref{fig:second_der}. It can be seen that the interpolated second derivative oscillates around the smooth exact second derivative. Using this interpolated data, the evolution produces an energy error (with respect to the exact solution) of $2.8\times 10^{-5}$ Hartree for the geometry change of $0.25$ $a.u.$ to $0.5$ $a.u.$. This shows that measuring a small number of points with high precision can provide accurate results. 

\subsubsection{Using many noisy measurements \label{sec:many_noisy}}
Another method that could be used to provide accurate reconstructed potentials and energies is to sample many time points with less precision and then smooth the data to reconstruct a density evolution. The $75$ time points generated from the discrete-clock simulation do not provide enough data to provide reasonable results using this method. Therefore, we return to the ground state of the system of Eq. \ref{eq:lih_system} with the density sampled every $0.012$ time interval as done in section \ref{sec:exact}. To approximate sampling error, we generate new noisy data $\tilde{n(t)}$, by taking the exact density evolution data $n_i(t)$ for point $i$ at time $(t)$ from section \ref{sec:exact} and apply a random Gaussian noise approximating ten thousand measurements per point defined as $\tilde{n_i(t)}=n_i(t)+G(1)*\sqrt{\frac{\left(1-n_i(t)\right)n_i(t)}{10000}}$ where $G(1)$ is a normal distribution with $\sigma=1$. The total data set now includes $12000$ noisy samples for time $144$. To generate this data on a quantum computer, one could either use the higher order discrete clock method to more distinct times, or use a lower order method and sample after each time step. The first-order Trotter method has good scaling for adiabatic evolutions\cite{Kovalsky2023} and would be useful in this case of frequent sampling. After generating the noisy densities, a locally weighted scatterplot smoothing (LOWESS) filter\cite{Cleveland1979} is used to average over one thousand time steps. A 4th order B-Spline is then interpolated using values every $300$ time steps of the smoothed data with the derivatives taken from the spline for all intermediate time steps. 

\begin{figure}
     \centering
     \begin{subfigure}[b]{0.45\textwidth}
         \centering
         \includegraphics[width=\textwidth]{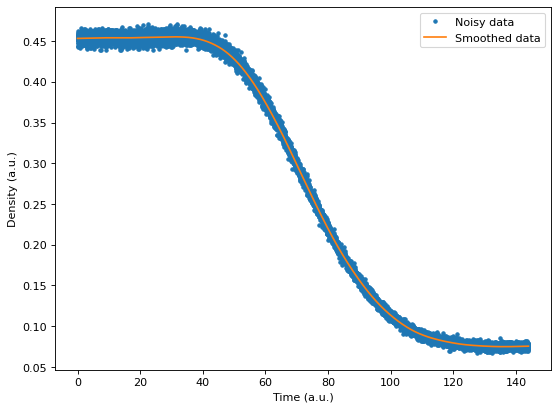}
         \caption{Smoothing LiH noisy data}
         \label{fig:smoothing}
     \end{subfigure}
     \hfill
     \begin{subfigure}[b]{0.45\textwidth}
         \centering
         \includegraphics[width=\textwidth]{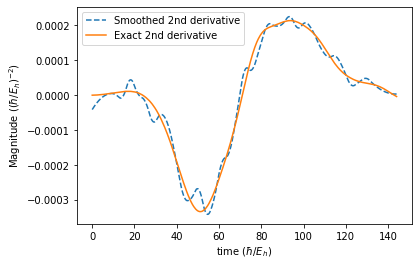}
         \caption{LiH 2nd time-derivative of density }
         \label{fig:smooth_v_exact}
     \end{subfigure}
        \caption{a) The noisy data with the smoothed time evolution of the density and the b) second time-derivative of the density error of the adiabatic evolution for the singlet state using smooth data after adding noise representing 10,000 samples.}
        \label{fig:Lih_singlet_noise_2nd}
\end{figure}

To approximate a noisy density matrix measurement, we use the smoothed density at $t=0$ for the diagonal elements of the required 1-RDM and then apply noise to the off-diagonal elements as $\tilde{\rho_{ij}(t)}=\rho_{ij}(t)+G(1)*\sqrt{\frac{\left|\left(1-\rho_{ij}(t)\right)\rho_{ij}(t)\right|}{10000}}$. The initial orbitals are generated using Eq \ref{eq.gen_orbs} with the noisy $\tilde{\rho_{ij}}$.  We then apply the algorithm using the noisy smoothed density data for $144 a.u.$ with the orbitals generated using the noisy RDM. The noisy data with the smoothed density evolution is shown in figure \ref{fig:smoothing} with the resulting second derivative compared to the exact second derivative of the density shown in figure \ref{fig:smooth_v_exact}. It appears from these figures that smoothing is a viable technique to recover the density evolution from many noisy measurements. However, it is most important to evaluate how well the smoothed evolution can be used to recover potential energy surfaces.

\begin{figure}
     \centering
     \begin{subfigure}[b]{0.45\textwidth}
         \centering
         \includegraphics[width=\textwidth]{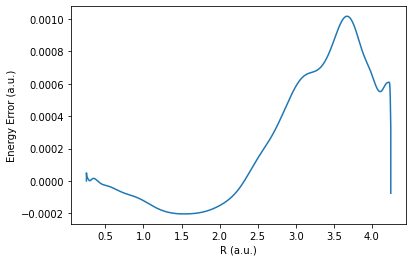}
         \caption{LiH energy error}
         \label{fig:e_err_noise}
     \end{subfigure}
     \hfill
     \begin{subfigure}[b]{0.45\textwidth}
         \centering
         \includegraphics[width=\textwidth]{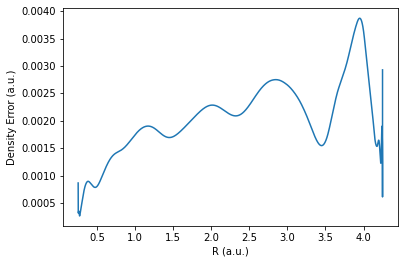}
         \caption{LiH density Error}
         \label{fig:d_err_noise}
     \end{subfigure}
        \caption{The a) energy and b) density error of the adiabatic evolution for the singlet state using smooth data after adding noise representing 10,000 samples.}
        \label{fig:Lih_singlet_noise}
\end{figure}

As can be seen from figure \ref{fig:Lih_singlet_noise}, the energy error is within chemical accuracy during the full time evolution and the density error is of the same magnitude as with the exact data. This shows that one can also sample noisily at many times and obtain reasonable accuracy for the full potential energy curve.

It should be noted that we did not attempt to utilize any physical constraints on the data during the smoothing process. The first property that could be utilized is that the density at every time should sum to the number of electrons. The second property results from the fact that we are using grid-like basis functions. Therefore, the density should be smooth in real space for our example system. In real molecules, there is a cusp (with known properties\cite{Kato1957}) in the density at the positions of nuclei. However, these positions are known and can be accounted for\cite{Kanungo2019}.

\subsection{Larger number of electrons}
To verify that the method works for more electrons, we utilize a Hamiltonian which requires fewer grid points for computational efficiency. This system is described by the potential
\begin{equation}
    V(t) = \sum_{i=1}^\eta\omega(t)r_i^2+\frac{1}{2}\sum_{i,j}^\eta\frac{1}{\sqrt{(r_1-r_2)^2+4}}
\end{equation}
where $\eta$ is the number of electrons and for all cases $\omega(t)$ goes from $1\rightarrow1.1$. The grid spacing required for convergence is $0.7 a_0$ but only $12$ grid points are required instead of $32$ due to the confinement of the harmonic potential. The exact evolution of the full wavefunction is performed in second quantization, here with the unrestricted Kohn-Sham\cite{Jacob2012} formalism used for the potential inversion, which means that the $\alpha$ and $\beta$ electrons can have different densities and Kohn-Sham potentials. For these tests, $\mathrm{d}t =0.012 $ for $5000$ time steps are taken for a total evolution time of $60$. For the ground state of a three-electron system with two $\alpha$ and one $\beta$ electrons, the energy error is $2.85 \times 10^{-7}$ Hartree with the $S^2$ value of $0.75007$ indicating some spin contamination. The line integral contribution for the three-electron system is $0.008$ Hartree. For a four-electron system with two $\alpha$ and two $\beta$ electrons, the energy error for the ground state is $1.5 \times 10^{-6}$ with the line integral contribution being $0.01765$ Hartree. The $S^2$ value for the 4-electron system was within numerical precision. The accuracy of these three and four electron systems shows that the algorithm is applicable to a higher number of electrons. The energy error for the 4-electron system is an order of magnitude larger than with 3-electron system but this could be coming from a larger adiabatic error. Further study is required to estimate the scaling with number of electrons but both errors are orders of magnitude less than chemical accuracy.

\subsection{Using Gaussian basis sets}
Since many basis functions are required to solve the electronic structure problem with grid-like bases, we turn to Gaussian basis sets to show that using the line integration technique can work for real molecules when the 1-RDM is determined along a path. The example system we study is the symmetric stretch of linear H$_4$ in a 6-31G(d) and G-31G(d,p) basis. The density we use is the full configuration interaction (FCI) 1-RDM along a path of geometries calculated using the PySCF integration in Tangelo. To obtain a reasonable approximation to the derivative of the density with respect to $R$, we obtain the 1-RDM from the FCI calculation at spacing of $0.025$ \AA between $[0.5, 2]$ \AA. Instead of using the TDDFT inversion method of reference \citenum{Brown2020}, we invert the potential directly with n2v\cite{n2v} using the Wu-Yang method with the Fermi-Amaldi guide potential and a regularization parameter of $0.001$ with a aug-cc-pVDZ potential basis set. We turn to n2v to avoid the spurious oscillations\cite{Gaiduk2013} that occur when attempting to invert the potential on a grid when the density comes from Gaussian basis sets.\cite{Yuming2021} $0.001$ is a large regularization parameter but is necessary due to the difficulty of the potential inversion process for this system. The derivative of the density with respect to $R$ is taken by transforming the measured 1-RDM to a $81\times 51\times 51$ grid in a box with sides of length $10$\AA. As the integration of the density is consistently about $4.001$, we renormalize the measured density to the correct $4$ (i.e. number of electrons) before numerically taken the derivative using a five-point stencil. Eq. \ref{eq.line_int} can then be utilized to obtain the potential energy curve.

\begin{figure}
     \centering
     \begin{subfigure}[b]{0.45\textwidth}
         \centering
         \includegraphics[width=\textwidth]{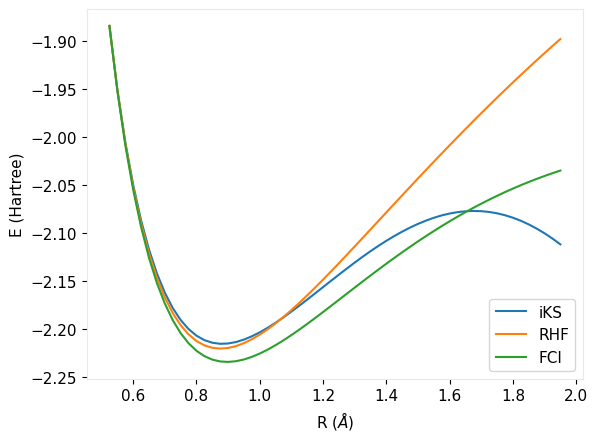}
         \caption{6-31G(d)}
         \label{fig:631gd_err}
     \end{subfigure}
     \hfill
     \begin{subfigure}[b]{0.45\textwidth}
         \centering
         \includegraphics[width=\textwidth]{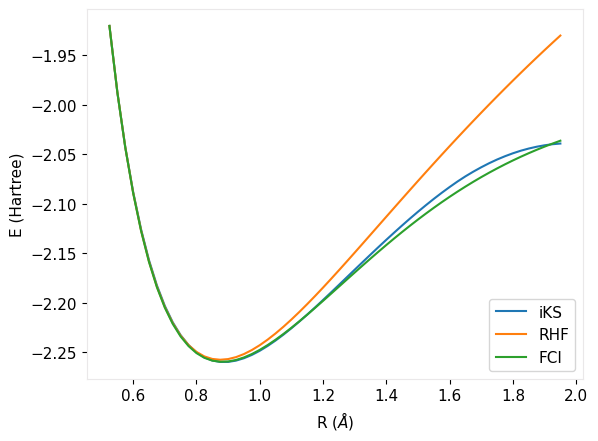}
         \caption{6-31G(d,p)}
         \label{fig:g31gdp_err}
     \end{subfigure}
        \caption{The potential energy curve of the linear H$_4$ with bond distance $R$  with a (a)6-31G(d) and (b)6-31G(d,p) basis. The curves are set such that the points are equivalent at $R=0.5$\AA. As can be seen, the curve calculated by inverting the potential (iKS) using the Wu-Yang method more closely follows the shape of the exact FCI curve than the Hartree-Fock (RHF) curve until the method fails at larger bond distances.}
        \label{fig:gauss_basis}
\end{figure}

As can be seen in Figure \ref{fig:gauss_basis}, utilizing incomplete basis sets such as Gaussian type orbitals (GTO) can provide reasonably accurate results but requires care that is not necessary with grid-like bases, such as regularization\cite{n2v, Yuming2021}. The curves are set such that the restricted Hartree-Fock (RHF), inverted Kohn-Sham (iKS) using Eq. \ref{eq.line_int} and full configuration interaction (FCI) values are all equivalent at 0.5\AA. The inversion process used here is non-linear and can often fail, and the large regularization parameter of $0.001$ means that we did not recover the ``exact'' KS potential as we had in the previous sections. At larger bond distances, the method starts to fail due to problems inverting the KS potential stemming from the inaccurate 1-RDM utilized. As expected, this failure occurs earlier with the smaller 6-31G(d) basis compared to the larger 6-31G(d,p) basis.

Increasing the basis further, and especially adding more diffuse basis functions, will increase the accuracy of the density at stretched geometries and permit the line-integration to succeed for larger bond distances\cite{n2v}. That being said, this does show that it is possible to use standard (GTO) basis sets and obtain reasonable results, especially near the equilibrium geometry. However, it should be noted that the Wu-Yang method requires the full 1-RDM at each geometry to perform the inversion. This is in opposition to the grid basis utilized in previous sections where only the density (diagonal of the 1-RDM) is required. One also does not have the ability to compare the density from the inverted potential to the measured density as n2v assumes that the measured 1-RDM corresponds to a ground state.

\section{Discussion and Conclusions}

\subsection{Generating full potential energy surfaces\label{sec:pes_gen}}
Due to the exponential scaling of the possible configurations of larger molecules, it is infeasible to scan all points on the surface. Therefore, it is necessary to be judicious about which geometries the adiabatic evolution samples. Two obvious candidate techniques can be used to represent the full potential energy of interest. 1) a sparse representation can be utilized to choose the necessary points to measure or 2) machine learning can be employed to model the density functional, to conduct efficient DFT calculations for non-measured geometries.

\subsubsection{Sparse representations}
It has been shown that it is not necessary to have a general surface representation of a system to obtain accurate vibrational energies of systems\cite{Wodraszka2021}. By only exploring fewer carefully-chosen points, an accurate description of the full system quantum dynamics can be calculated. One can also generate accurate potential energy surfaces using Smoylak grid points\cite{Avila2020} or machine learning the potential surface directly\cite{Manzhos2021}.

For larger systems, one may also be able to use a Reaction Path Hamiltonian\cite{Miller1980, Taketsugu1996} (RPH) if a known reaction path (or surface\cite{Carrington1984}) is known to be most important. A lower level of theory may then be used to calculate Harmonic frequencies along this reaction path with the accurate energies from the quantum simulation used to specify the energy along the reaction path. It may also be possible to utilize a  density functional modeled with machine learning to calculate the Hessian required for the RPH.

\subsubsection{Machine Learning\label{sec:ml}}
For systems where generating a full potential energy surface is unfeasible, one can attempt to recover the underlying density functional that represents the chemical system. It has been shown\cite{Li2021, Ryczko2022} that only a small number of points may be necessary to generate a density functional modelled with machine learning that describes the full potential energy surface. During the adiabatic evolution, TDDFTinversion approximates the first time-derivative of the potential to increase the accuracy of the inversion process. This means that we have access to $v_{KS}(r, t), v^{\prime}_{KS}(r,t), \rho(r, t), \rho^{\prime}(r, t), \rho^{\prime \prime}(r, t)$, and $v^{\prime}_{ext}(r, t)$ where $\prime$ signifies the time-derivative, as well as the full density profile on a grid to calculate gradients. This additional information should be useful in developing accurate functionals. In fact, it is known that the transferability of a density functional generated with machine learning is improved when training on both the density and energy as opposed to just the energy\cite{Li2021, Bai2022}. It may also be useful to calculate multi-state machine-learned functionals\cite{Bai2022} as we have shown that excited states can be calculated using the method outlined here.

\subsection{Comparison with QPE\label{sec:vqpe}}
If one is already interested in the density at a variety of geometries to calculate properties such as forces acting within molecules\cite{Feynman1939} or utilize as a fidelity witness\cite{Skogh2024} by comparing with experiment, the extra evolution time required for QPE is unnecessary as all information to calculate potential energy surfaces is already present.

To compare the cases when the technique outlined in this manuscript can provide a more efficient technique to obtain only the energy information, we look to the performance bounds of QPE\cite{Reiher2017} and adiabatic evolution\cite{Kovalsky2023}. In this comparison, we will stick to first-order Trotterization as bounds have been presented in literature for both time-independent\cite{Reiher2017} and time-dependant adiabatic\cite{Kovalsky2023} evolution. For both cases, the Trotter formula only converges when $\mathrm{d}t \big|\big|H(t)\big|\big|\in \Omega\left(1\right)$\cite{Reiher2017, Kovalsky2023} where $\mathrm{d}t$ is the time-step.

\subsubsection{QPE Trotter steps}
For QPE, the total time-evolution required is $T > \frac{1}{\epsilon}$ where $\epsilon$ is the required accuracy. The Trotterization error for the energy of the approximate unitary can be bounded by\cite{Reiher2017}
\begin{equation}
    E-E_{t} < \alpha \,\mathrm{d}t
\end{equation}
where $\alpha = 2\sum_m \big|\big| \left[H_m, \sum_{m^{\prime}<m} H_{m^{\prime}}\right]\big|\big|$ and $\left[H_i, H_j\right]$ is the commutator between terms $H_i$ and $H_j$ in the Hamiltonian. Therefore, the total number of Trotter-steps required $\mathcal{N}$ is
\begin{equation}\label{eq.qpeN}
    \mathcal{N} = T\,\mathrm{d}t^{-1}=\frac{1}{\epsilon}\frac{\alpha}{\epsilon}=\frac{\alpha}{\epsilon^2}
\end{equation}
Therefore,  QPE requires Trotter steps on the order of $\mathcal{O}\left(\epsilon^{-2}\right)$ to estimate an energy of the Hamiltonian with error $\epsilon$. The probability of success is proportional to $1/c^2$ where $c$ is the overlap of the desired state with the initial state. To obtain a single point energy therefore requires $\mathcal{O}\left(\epsilon^{-2}c^{-2}\right)$ Trotter steps.

\subsubsection{Adiabatic state preparation Trotter steps.}
For the algorithm presented here, we require that $c \approx 1$ which we utilize adiabatic state preparation to achieve. For the time-dependent adiabatic error, the scaling is much better when the state is measured at the end-point of the evolution due to the self-healing of Trotter errors\cite{Kovalsky2023}. One could do this by using the scheduling function of Eq \ref{eq.sched} but with the end point changed to different $R$. We simply need a prepared state at each geometry and the adiabatic KS system can still act as a witness of the success of these separate state preparations. The infidelity $\epsilon_I$ of the adiabatic evolution is\cite{Kovalsky2023}
\begin{equation}\label{eq:ei}
    \epsilon_I=1-\bra{\psi}U(\infty)^{\dagger}U(T)\ket{\psi} < \mathcal{O}\left(\mathrm{d}t^2T^{-2}\right)+\mathcal{O}\left(C_1\frac{1}{T^2}\right),
\end{equation}
where $C_1$ depends on the energy gap $\Delta$ to the nearest state and the speed at which the Hamiltonian is changing during the evolution. The amount of time-evolution required to exclude diabatic transitions is such that $T>\frac{\sum_m \big|\big|\dot{H}(t) \big|\big|}{\Delta^2}$ where $\dot{H}(t)$ is the first-derivative of the Hamiltonian with respect to $t$. The number of Trotter steps required to adiabatically evolve a system into the desired state once is
\begin{equation}
    \mathcal{N}>T\, \mathrm{d}t^{-1} = T \mathcal{O}\left(\frac{\sqrt{\epsilon_I}}{T}\right)= \mathcal{O}{\frac{1}{\sqrt{\epsilon_I}}}
\end{equation}
where the $\mathrm{d}t =\mathcal{O}\left(\frac{1}{\sqrt{\epsilon_I}T}\right)$ comes from the first term in Eq \ref{eq:ei}. If we assume that the infidelity in the state $\epsilon_I$ corresponds to the error in the measured density ($\epsilon_d$) then the number of Trotter steps required to obtain $\epsilon_d$ is given by $\mathcal{N} = \mathcal{O}\left(\epsilon_d^{-1}\right)$. This is the main result of reference \citenum{Kovalsky2023} suggesting that the success of adiabatic state preparation can be increased solely by increasing the number of Trotter steps. We do need to use a valid $\mathrm{d}t$ (i.e. $\mathrm{d}t \big|\big|H(t)\big|\big|\in \Omega\left(1\right)$), therefore, the proper bounds to prepare a state are
\begin{equation}\label{eq.adiaN}
    \mathcal{N}> \mbox{max}\left(\mathcal{O}\left(\frac{1}{\epsilon_d}\right), \big|\big|H(t)\big|\big|\frac{\sum_m \big|\big|\dot{H}(t) \big|\big|}{\Delta^2}\right)
\end{equation}

Comparing Eq. \ref{eq.qpeN} and Eq. \ref{eq.adiaN} reveals that the total number of Trotter steps could be much lower to prepare a state using adiabatic state preparation as opposed to using a full QPE calculation (which also prepares the state). This difference depends on the energy gap, the norm of the Hamiltonian, and the commutator error during the adiabatic evolution. However, as discussed in the introduction, preparing the state is not sufficient to obtain the energy.

Obtaining bounds on the energies calculated from the techniques used in this manuscript is difficult due to the multiple heuristics used. A full analysis will be left to future work but we can discuss some approximate asymptotic scaling. In general, sampling observables scales as $\mathcal{O}\left(1/\epsilon_d^2\right)$. However, one could measure the $N$ density observables by adding $\mathcal{O}\left(N\log(1/\epsilon) \right)$ more qubits and utilizing a circuit that prepares (or unprepares) the state $\mathcal{O}\left(\sqrt{N}/\epsilon\right)$\cite{Babush2023, Huggins2022} times in succession. As $N$ will be very large, this could be an issue but fewer observables could be measured simultaneously at the cost of more repetitions.\cite{Huggins2022} It is difficult to know the relationship between $\epsilon_d$ and $\epsilon$, but for this simplified analysis, we will assume they are related through some constant. Therefore, the algorithm presented here will require 
\begin{equation}\label{eq.n}
\begin{split}
    \mathcal{N}> \mbox{max}\left(\mathcal{O}\left(\frac{1}{\epsilon^3}\right), \mathcal{O}\left(\big|\big|H(t)\big|\big|\frac{\sum_m \big|\big|\dot{H}(t) \big|\big|}{\Delta^2\epsilon^2}\right)\right) \, \mbox{or} \, \\ \mbox{max}\left(\mathcal{O}\left(\frac{\sqrt{N}}{\epsilon^2}\right), \mathcal{O}\left(\sqrt{N}\big|\big|H(t)\big|\big|\frac{\sum_m \big|\big|\dot{H}(t) \big|\big|}{\Delta^2\epsilon}\right)\right),
\end{split}
\end{equation}
Trotter steps to obtain the density at a single geometry to error $\epsilon$.

\subsubsection{Regime of algorithm advantage}
As we are obtaining energy evaluations at many points, one would also need to rerun QPE multiple times as a comparison to obtain a good representation of the potential energy surface. For QPE, one could obtain a good representation of the surface using a similar number of points as in section \ref{sec:few_exact} which was $75$ for our use case. In this few accurate measurement regime, one could perform $\mathcal{O}\left(1/\epsilon^2\right)$ repetitions at each point for the density to have a direct comparison when sampling the density or use the larger circuit of reference \citenum{Huggins2022} once.

Putting this all together, Table \ref{tab:my_label} shows the comparison in Trotter steps, repetitions and ancilla qubits. It is clear that the total space-time complexity of this algorithm is unlikely to be better than QPE  asymptotically for measuring the energy alone. The only regime where we can definitively say that this technique could prove beneficial (for measuring the energy alone) is in the early fault tolerant era where long QPE circuits are not possible due to finite error correction.\cite{Liang2023, Zhang2022} 

\begin{table}[]
    \centering
    \begin{tabular}{|c|c|c|c|}
    \hline
        Method & Trotter steps & Repetitions & Ancilla Qubits\\
        \hline
        Here + sampling &  $\mathcal{O}\left(\frac{1}{\epsilon}\right)$ & $\mathcal{O}\left(\frac{1}{\epsilon^2}\right)$ & $0$ \\
        Here +  Ref. \citenum{Huggins2022} &  $\mathcal{O}\left(\frac{\sqrt{N}}{\epsilon^2}\right)$ & $1$ & $\mathcal{O}\left(N\log(\frac{1}{\epsilon})\right)$ \\
        QPE &  $\mathcal{O}\left(\frac{1}{\epsilon^2}\right)$ & $\mathcal{O}\left(\frac{1}{c^2}\right)$ & $\mathcal{O}\left(\log(\frac{1}{\epsilon})\right)$ \\
        adiabatic + QPE &  $\mathcal{O}\left(\frac{1}{\epsilon}+\frac{1}{\epsilon^2}\right)$ & 1 & $\mathcal{O}\left(\log(\frac{1}{\epsilon})\right)$ \\
        Ref. \citenum{Lin2020} + QPE &  $\mathcal{O}\left(\log(\frac{1}{\epsilon})+\frac{1}{\epsilon^2}\right)$ & 1 & $\mathcal{O}\left(\log(\frac{1}{\epsilon})\right)$ \\
        \hline
    \end{tabular}
    \caption{The comparison of the method introduced in this paper vs  QPE for a single accurate energy/density measurement. The above circuits would need to be repeated multiple times to generate a good representation of the PES. }
    \label{tab:my_label}
\end{table}

That being said, the Hohenberg-Kohn theorem\cite{HohenbergKohn} states that there is a one-to-one mapping between the external potential and the density. Therefore, in principle, all properties of a system can be determined from the density alone.\cite{Skogh2024} Here, we showed how energy differences could be obtained from the density but it is certainly possible that other properties could be obtained using the density information gathered. One possible route to this is discussed in section \ref{sec:ml} and ref \citenum{Li2021}. The density could be used to improve classical quantum chemistry, and especially DFT approximations for the system of interest. This could allow many additional geometries and properties to be calculated more accurately without accessing a quantum computer.

\subsection{Closing remarks\label{sec:conclude}}
We have evaluated the applicability of using DFT to calculate energies from quantum computers that possibly does not require the long-time evolution of QPE and requires far fewer measurements than variational quantum eigensolvers VQE\cite{Gonthier2022}. Instead of performing DFT on the quantum computer directly\cite{Senjean2023, Ko2023implementation}, we proposed evaluating the DFT path integral to generate potential energy surfaces. The path integral was evaluated using TDDFT potential inversion along an adiabatic time evolution between different molecular geometries by only using the measured density. We have shown that the method can obtain chemical accuracy  along a path of external potentials for two- three- and four-electron systems. Chemical accuracy is retained if either few time-points are measured accurately or many time points are measured inaccurately. Regarding the latter, LOWESS smoothing reproduces the exact density evolution well enough for the algorithm present here to succeed.

In the future, it will be necessary to see if this method can be applied to systems where adiabatic time evolution may require very-long times due to small energy gaps between different potential energy surfaces. The difference in the measured density and the calculated density from the inverted potential appears to provide a reasonable proxy for the applicability of the adiabatic approximation and the accuracy of the calculated energy. As digital adiabatic evolution can be improved by increasing the time without decreasing the time-step size\cite{Kovalsky2023}, the technique should be systematically improvable via a single parameter. More advanced quantum control techniques could also be applied\cite{Li2023}. 

Although our focus has been on calculating single state energy levels, we can remark on the paths one may take to obtain excitation energies between states. If, as we did for the LiH model system, calculate multiple states surfaces at the same geometry then it may be possible to utilize ensemble DFT\cite{Gould2022} to calculate excitation energies. As the method of reference \citenum{Brown2020} generates the exact TDDFT potential, one could also use real-time TDDFT to calculate excitation energies. Although in this case, it may be better to directly perform the Fourier transform on the measured densities\cite{Chan2023b}, after applying a perturbing potential, but this will have the same $\mathcal{O}(1/\epsilon)$ evolution time requirements as QPE. One could also use the modified Sternheimer method\cite{Yabana2006} using the inverted $V_{KS}$ potential but this will generally not be chemically accurate.

\section{Acknowledgements}
The author would like to thank Alexandre Fleury, Erika Lloyd, Valentin Senicourt and Marc Coons for suggestions to improve the research and manuscript.


\section{Appendix}
\subsection{Avoided crossing\label{sec:avoided}}
\begin{figure}
     \centering
     \begin{subfigure}[b]{0.45\textwidth}
         \centering
         \includegraphics[width=\textwidth]{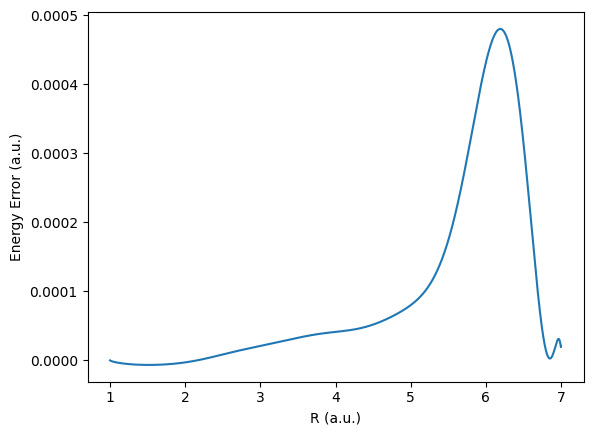}
         \caption{LiH energy error}
         \label{fig:e_err_bigc}
     \end{subfigure}
     \hfill
     \begin{subfigure}[b]{0.45\textwidth}
         \centering
         \includegraphics[width=\textwidth]{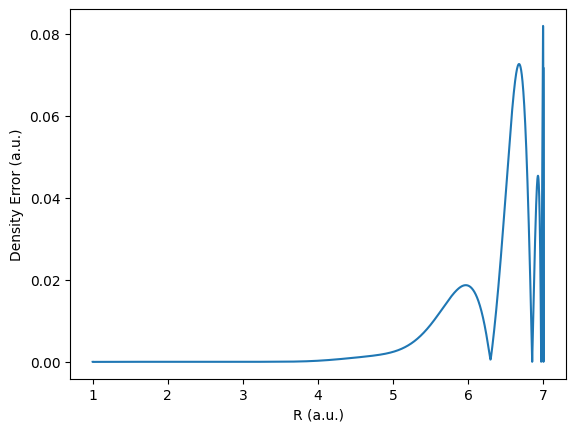}
         \caption{LiH density Error}
         \label{fig:d_err_bigc}
     \end{subfigure}
        \caption{The a) energy and b) density error of the adiabatic evolution for the ground singlet state with $c=2.8$. The evolution time of the KS system is still too short due to the small energy difference between the two lowest KS states at this stretched geometry. However, the energy error is still small.}
        \label{fig:Lih_singlet_bigc}
\end{figure}

When using $c=2.8$ to define the system of Eq \ref{eq:lih_system}, the two lowest singlet state potential energy curves have a much closer avoided crossing as can be seen in figure \ref{fig:c28} compared to figure \ref{fig:c06} with $c=0.6$. We can apply the technique but the evolution time required to obtain an overlap at the end of $0.999$ is much longer being $864$ a.u. for an evolution from $R=1$ to $R=7$ Bohr. For the KS system to remain adiabatic in the ground state requires even longer evolution time. We used a KS time evolution of $1152$ a.u. and as can be seen in figure \ref{fig:d_err_bigc}, this is still too short of time. However, the energy error (figure \ref{fig:e_err_bigc}) is still well within chemical accuracy.

\subsection{Discrete Clock applied to grid-like basis functions\label{app:dc_grid}}
The discrete-clock construction\cite{Watkins2022} is a method that approximates the time-evolution of a system as a linear combination of $M$ different $k$-step second-order Trotter evolution unitaries. The evolution of the Hamiltonian at time $t$ for time-step $\mathrm{d}t$  is approximated as
\begin{equation}
    \exp\left\lbrace -iH(t+\mathrm{d}t/2)\mathrm{d}t\right\rbrace \approx U_{M}\left(t+\mathrm{d}t, t\right):=\sum_{j=1}^M a_j U_2^{(k_j)}\left(t+\mathrm{d}t,t\right)
\end{equation}
where $a_j$ are the coefficients for each of the $k_j$-step second-order Trotter $U_2^{(k_j)}$
\begin{equation}
    U_{2}^{(k_j)}\left(t_0+\mathrm{d}t, t_{0}\right) := \prod_{q=1}^{k_j} U_2\left(t_0 + \mathrm{d}t \,q/k_j, t_0+\mathrm{d}t\left(q-1\right)/k_j\right)
\end{equation}

The test system we use to test the discrete clock method is first presented in reference \citenum{vanDijk2014} which for the time-dependant Hamiltonian
\begin{equation}
    H(t) = \frac{\mathrm{d}^2}{\mathrm{d} x^2}+\left(4e^{-2t}-\frac{1}{16}\right)x^2-2e^{-t}
\end{equation}
which has the exact solution
\begin{equation}
    \psi\left(x,t\right) = \left(\frac{2}{\pi}\right)^{1/4}\exp\left(-x^2e^{-t}-\frac{1}{4}t+i\frac{x^2}{8}\right)
\end{equation}

\begin{figure}
    \centering
    \includegraphics[width=0.45\textwidth]{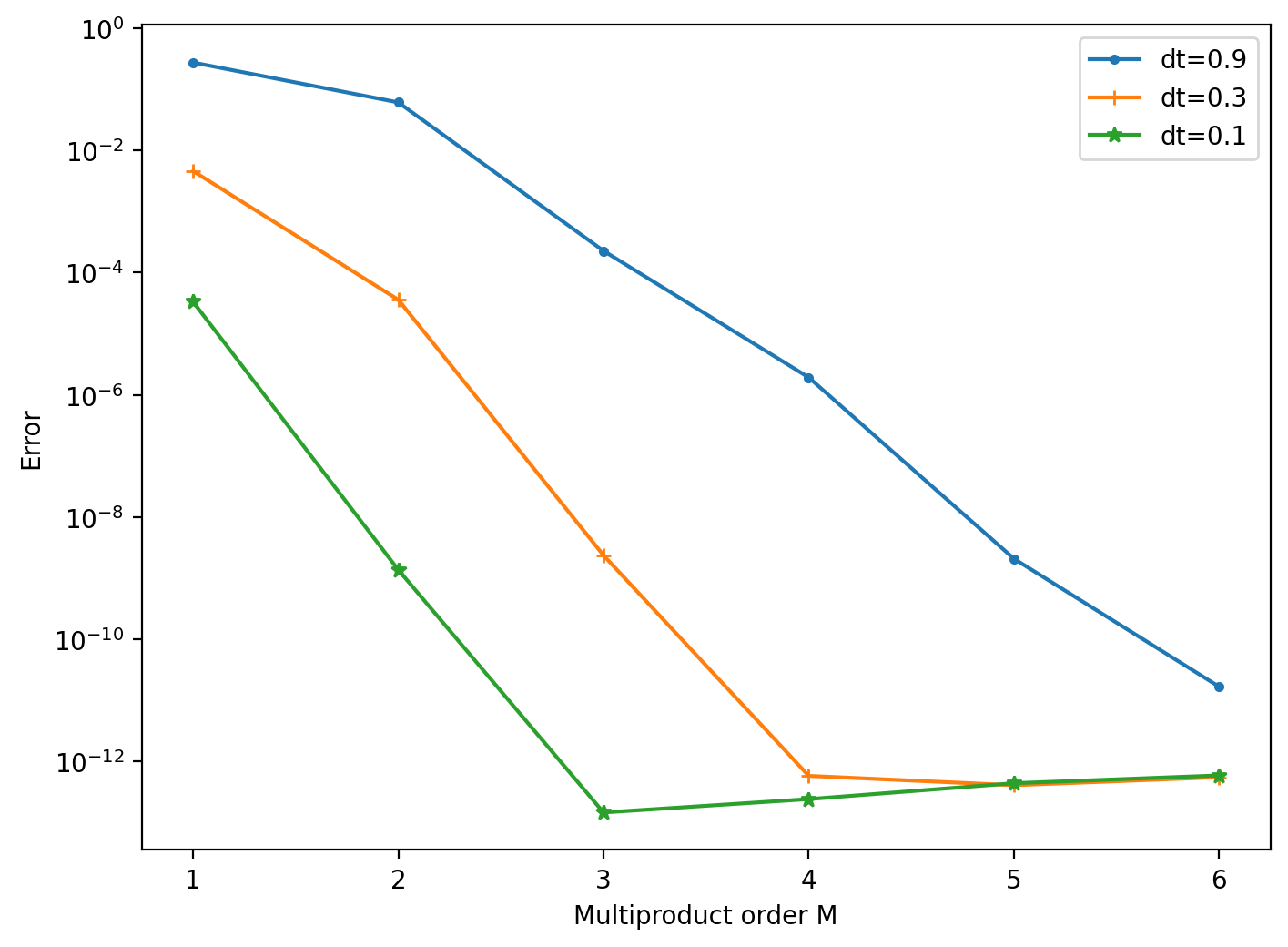}
    \caption{The error of the time evolution for a time step $\mathrm{d}t$ and given order multiproduct order $M$.}
    \label{fig:discrete_clock}
\end{figure}

As can be seen in figure \ref{fig:discrete_clock}, the error decreases rapidly with increasing multiproduct order. However, the number of time-steps included corresponds to the accuracy in this case. $M=1$ requires one second-order Trotter, $M=2$ requires $1+2=3$ second-order Trotter time steps and $M=3$ requires $1+2+6=9$ second-order Trotter time-steps. However, the unique times at which the potential is evaluated are $1,3$ and $7$ respectively due to the time steps for $2$ time-step Trotter being encompassed in the $6$ time-step evolution.

This results in a simplification of the circuit being possible as shown in figure \ref{fig:2_6_comb} where the CNOT ladders for the $2$ time-steps are merged into the $6$ time-steps. This circuit assumes that the state is on $q_1, q_2, q_3$ and the coefficients for the linear combination of unitaries decomposition are stored in $\alpha_1, \alpha_2, \alpha_3$ as the vector $=[c_f/2,c_f/2,0,0,c_1,c_2,c_6]$. Where $c_f$ is used such that $c_1+c_2+c_6+c_f=2$ for oblivious amplitude amplification and the multiproduct formula is defined $\sum_{i=1}^{3} c_i \prod_{i=1}^{k_i} U(t_{k,i})$.

\begin{figure}
    \centering
\begin{quantikz}
\lstick{$\ket{\alpha_3}$} & \qw & \qw & \qw & \octrl{5} & \qw &\qw & \qw \\
\lstick{$\ket{\alpha_2}$} & \qw & \qw &\ctrl{4} &\ctrl{4} & \qw&\qw & \qw \\
\lstick{$\ket{\alpha_1}$} & \qw & \qw &\ctrl{3} &\ctrl{3} & \qw &\qw &\qw \\
\lstick{$\ket{q_3}$} & \ctrl{1} & \qw  & \qw& \qw&\qw & \ctrl{1} &\qw \\
\lstick{$\ket{q_2}$} & \gate{X} & \ctrl{1}  & \qw & \qw&\ctrl{1} \qw & \gate{X}& \qw \\
\lstick{$\ket{q_1}$} & \qw & \gate{X}  & \gate{R_z(2c \,\mathrm{d}t/6)}& \gate{R_z(4c\, \mathrm{d}t/6)}&\gate{X} &\qw&\qw
\end{quantikz},
\caption{CNOT ladder simplification for combining the $2$ and $6$ step multiproduct evolution for $\exp(-ic Z_1 Z_2 Z_3 \mathrm{d}t)$}
\label{fig:2_6_comb}
\end{figure}

\bibliography{references}

\newpage
\begin{figure}
    \centering
    \includegraphics[width=3in]{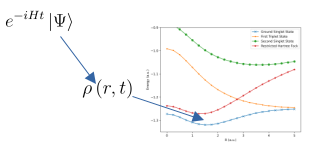}
    \caption{TOC graphic}
\end{figure}

\end{document}